\documentclass[intlimits,twoside,a4paper]{article}
\usepackage{amsmath,amssymb,gensymb}

\usepackage[cp1251]{inputenc}

\usepackage{cmpj3}

\issue{2017}{20}{4}{43702}
\doinumber{10.5488/CMP.20.43702}
\title[Magnetoresistance based determination of basic parameters of minority charge carriers]%
{Magnetoresistance based determination of basic parameters of minority charge carriers in solid matter}
\author[Y.O.~Uhryn, R.M.~Peleshchak, V.B.~Brytan, A.A. Velchenko]{Y.O.~Uhryn\refaddr{label1}\thanks{E-mail: yuriyuhryn@yahoo.com.ua}\,, R.M.~Peleshchak\refaddr{label1}, V.B.~Brytan\refaddr{label1}, A.A. Velchenko\refaddr{label2}}
\addresses{
\addr{label1} Drohobych Ivan Franko State Pedagogical University, 24 I. Franko St., 82100 Drohobych, Ukraine
\addr{label2} Belarusian State Agrarian Technical University, 99 Nezavisimosti Ave., 220023 Minsk, Belarus
}

\date{Received September 7, 2016, in final form May 31, 2017}
\begin{document}

\maketitle

\begin{abstract}
Magnetoresistance as a tool of basic parameters determination of minority charge carriers and the ratio of minority charge carriers conductivity to majority ones  in solid matter has been considered within the framework of the phenomenological two-band model.
The criterion of the application of this model has been found.
As examples of these equations usage the conductor, semiconductor and superconductor have been introduced. From the obtained temperature dependences of the aforementioned values in superconductor, a supposition of a deciding role of minority charge carriers in the emergence of superconductivity state has been made.

\pacs  72.15, 72.20
\end{abstract}

\section{Introduction}

Up till now, the researchers have found that magnetoresistance measurements are not informative concerning basic parameters of charge carriers  in solid matter \cite{Kuc}. Among the galvanomagnetic effects only the Hall-effect is presented as a tool for determining these parameters \cite{Kuc}. However, we are going to show that a magnetoresistance curve conceals in itself an exact information about minority charge carriers mobility and concentration as well as their conductivity in relation to the conductivity of the majority ones.

The aim of this paper is to develop a magnetoresistance method to determine basic parameters of the minority charge carriers in solid matter \cite{Uhr1} and apply it to  different solids.

\section{Calculation}

To achieve this aim, we must consider the phenomenological model of galvanomagnetic phenomena for an isotropic material with two types of charge carriers (having opposite or the same sign). It is known \cite{Smi1} that the transverse conductivity $\sigma $ for this case depends upon the magnetic field induction in the following way:

\begin{equation} \label{GrindEQ__1_} \sigma =\frac{(\sigma _{1} +\sigma _{2} )^{2} +\sigma _{1} ^{2} \sigma _{2} ^{2} (R_{\text{H1}} +R_{\text{H2}} )^{2} B^{2} }{(\sigma _{1} +\sigma _{2}^{} )+\sigma _{1} \sigma _{2} (\sigma _{1} R_{\text{H1}} ^{2} +\sigma _{2} R_{\text{H2}} ^{2} )B^{2} }\,.
\end{equation}
Substitute here $\sigma _{1}^{} =en\mu _{n}^{} $, $\sigma _{2} =ep\mu _{p} $, $R_{\text{H1}} =-{1 \mathord{\left/{\vphantom{1 (en)}}\right.\kern-\nulldelimiterspace} (en)}$, $R_{\text{H2}} ={1 \mathord{\left/{\vphantom{1 (ep)}}\right.\kern-\nulldelimiterspace} (ep)}$, where $e$ is electron charge, $n,\, \, p,\, \, \mu _{n} ,\, \, \mu _{p} $ are concentrations and mobilities of two types of charge carriers (it can be electrons and holes or light and heavy holes). Note also that $\sigma =\rho /(\rho ^{2} +\rho _{xy} ^{2} )$, where $\rho $ is transverse and $\rho _{xy} $ Hall resistivity. Then, considering that $\rho _{xy} \ll\rho$, we have
\begin{equation} \label{GrindEQ__2_} \rho =\frac{1}{\sigma } =\frac{1}{e} \frac{n\mu _{n} +p\mu _{p} +\mu _{n}^{} \mu _{p} (n\mu _{p} +p\mu _{n} )B^{2} }{(n\mu _{n} +p\mu _{p} )^{2} +\mu _{n} ^{2} \mu _{p} ^{2} (n-p)^{2} B^{2} }\,.
\end{equation}
If we state now the condition $\rd^{2} \rho /\rd B^{2} =0$, we obtain the magnetic field position of $\rho $ flex point $B_{\text f} $
\begin{equation} \label{GrindEQ__3_} B_{\text f} =\frac{1+ab}{\sqrt{3} (1-a)\mu _{n} }\,,
\end{equation}
where $a=n/p$, $b=\mu _{n} /\mu _{p} $.

Now, the equation (\ref{GrindEQ__2_}) can be introduced by the form:
\begin{equation} \label{GrindEQ__4} \rho =\rho _{0} \frac{1+\frac{\mu _{n} (a+b)}{\sqrt{3} B_{\text f} (1-a)b} B^{2} }{1+\frac{1}{3B_{\text f} ^{2} } B^{2} }\,,
\end{equation}
where $\rho _{0} $ is resistivity at $B=0$.

Let us write the condition for electrons as minority charge carriers. It is $a\ll1$. Since the minority charge carriers mobility is, as the rule, higher or compared with the majority ones, that is $b\geqslant 1$, then $a\ll b$ and the magnetic field dependence of $\rho $ gets the form:
\begin{equation} \label{GrindEQ__5_} \rho =\rho _{0} \frac{1+\frac{\mu _{n} }{\sqrt{3} B_{\text f} } B^{2} }{1+\frac{1}{3B_{\text f} ^{2} } B^{2} }\,.
\end{equation}

 For $B=B_{\text f} $, writing for convenience $\rho (B_{\text f} )=\rho _{\text f} $, we obtain:
\begin{equation} \label{GrindEQ__6_} \mu _{n} =\frac{\sqrt{3} }{B_{\text f} } \left(\frac{4}{3} \frac{\rho _{\text f} }{\rho _{0} } -1\right).
\end{equation}

Thus, as can be seen from the last formula if the condition\textit{ $n\ll p$ }is fulfilled, the measuring of the transverse magnetoresistance provides the information about the mobility of minority charge carriers. It is enough to find the magnetoresistance flex point  $B_{\text f} $ and to measure the resistivity in this point $\rho _{\text f} $ and in zero field $\rho _{0} $.

From equation (\ref{GrindEQ__3_}) for $a\ll 1$, taking into account that $ab=\sigma _{n} /\sigma _{p} $, we get another useful formula
\begin{equation} \label{GrindEQ__7_} \frac{\sigma _{n} }{\sigma _{p} } =4\left(\frac{\rho _{\text f} }{\rho _{0} } -1\right).
\end{equation}

Since in formula~\eqref{GrindEQ__6_} and \eqref{GrindEQ__7_} the resistivities appeared in ratios, in practice we can substitute $\rho _{\text f} /\rho _{0}^{} $ by a corresponding ratio of potential differences $U_{\text f} /U_{0} $.

In the strong field limit, taking into account that $a\ll1$ and $a\ll b$, equation~(\ref{GrindEQ__2_}) results after the simple treatment in the formula for saturation resistivity:
\begin{equation} \label{GrindEQ__8_} \rho _{\infty } =\frac{1}{\sigma {}_{p} }\,.
\end{equation}

From this equation, equations~(\ref{GrindEQ__6_}) and (\ref{GrindEQ__7_}) and taking into account that $\sigma _{n} =en\mu _{n} $, we obtain a formula for determining the concentration of minority charge carriers based on the experimentally measured values
\begin{equation} \label{GrindEQ__9_} n=\frac{4B_{\text f} }{\sqrt{3} e\rho _{\infty } } \frac{\frac{\rho _{\text f} }{\rho _{0} } -1}{\frac{4}{3} \frac{\rho _{\text f} }{\rho _{0} } -1}\,.
\end{equation}

The latter formula presumes also a possibility to avoid measuring in high magnetic field for determining $n$, so equation~(\ref{GrindEQ__9_}) gets the form:
\begin{equation} \label{GrindEQ__12_} n=\frac{4\sqrt{3} B_{\text f} }{e\rho _{0} } \frac{\frac{\rho _{\text f} }{\rho _{0} } -1}{\left(4\frac{\rho _{\text f} }{\rho _{0} } -3\right)^{2} }\,.
\end{equation}

\begin{figure}[!b]
\centerline{\includegraphics[width=0.45\textwidth]{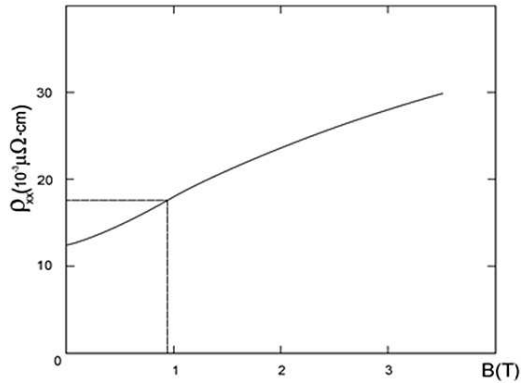}}
\caption{ Magnetic field dependence of transverse resistivity for refined Al. Dashed lines indicate the flex point $B_{\text f} $ and resistivity in this point $\rho \left(B_{\text f} \right)$. The curve is obtained from experimental data of \cite{Luc}. } \label{fig1}
\end{figure}

Let us return to the equation~(\ref{GrindEQ__5_}). In the strong field limit we can neglect 1 in comparison with the summand containing \textit{B}  in the numerator and denominator of (\ref{GrindEQ__5_}), getting another formula for determining~$\mu _{n} $
\begin{equation} \label{GrindEQ__13_} \mu _{n} =\frac{\rho _{\infty } }{\sqrt{3} \rho _{0} B_{\text f} }\,.
\end{equation}

Combining the latter formula with (\ref{GrindEQ__6_}) we get a relation between the experimentally measured values $\rho _{0}$, $\rho _{\text f} $ and $\rho _{\infty } $:
\begin{equation} \label{GrindEQ__14_} \rho _{\infty } =4\rho _{\text f} -3\rho _{0}\,,
\end{equation}
which is a test relation for the semi-classic behavior (\ref{GrindEQ__2_}) of a real experimental curve.

We shall illustrate some examples of using the introduced equations.

\section{Examples of calculation}

We have applied these results to the metal aluminum at $T=4.2$~K. As Al is located in the third group of the periodic table of elements, its atom has 3~valent electrons (3$s^23p^1$). The analysis of the Hall-effect experimental data \cite{Luc} proves that out of these three electrons only two become free, the third one can tunnel through the potential barrier to the neighboring atom due to the overlap of the wave functions. To this free place, an electron from another neighboring atom can come and so on. Thus, this free place moving in a chaotic manner within the crystal behaves as a positive free particle with electron charge --- the hole. The presence of a hole in Al is confirmed both by Fermi surface calculation and Hall-effect experiments \cite{Luc1}. Dispersion law of free charge carriers in the second Brillouin zone is pointed out by the hole nature of these carriers \cite{Luc1} and as in this zone there is approximately one of the three valent electrons, then the hole concentration must be approximately twice smaller than the free electron concentration \cite{Ash}. This is observed experimentally: the Hall coefficient depends on the magnetic field and changes its sign from negative to positive in a strong magnetic field, in which it is twice larger than in the weak one. Such a Hall-effect sign change is observed not only in Al, but at least in Be, Mg, In and Pb \cite{Ash}.

Figure~\ref{fig1} shows the magnetic field dependence of resistivity for a refined sample of this metal \cite{Luc}. In fact, the presence of a transverse magnetoresistivity is the evidence of the existence of two types of charge carriers. Calculations using formulae~(\ref{GrindEQ__6_})--(\ref{GrindEQ__9_}) for different refined Al are introduced in table~\ref{tab1}.

As it is seen from table~\ref{tab1}, the refining influences only  the mobility of charge carriers. Since the purification does not influence the ratio $\sigma _{n} /\sigma _{p}$, we can conclude that the hole mobility increases at the same rate as the electron mobility. 

\begin{table}[!t]
\begin{center}
\caption{Purification dependence of electron mobility $\mu _{n} $, electron concentration $n$, electron conductivity~$\sigma _{n} $ and hole conductivity $\sigma _{p}^{} $ for Al.}
\vspace{2ex}
\label{tab1}
\begin{tabular} {|l|c|c|c|c|c|} \hline\hline
Purification & $\mu _{n}$, $\frac{\text{m}^{2\strut}}{\text{V}\cdot \text{s}}$ & $n$, $10^{28}~\text{m}^{-3}$ & $\sigma _{n}$, $10^{9}~\text{\textohm} \cdot \text{m}$ & $\sigma _{p}$, $10^{9}~\text{\textohm} \cdot \text{m}$ & $\sigma _{n}/\sigma _{p} $ \\ \hline\hline
Refining & 1.75 & 1.9 & 5.4 & 2.9 & 1.87 \\ \hline
Zone refining & 14.3 & 2.0 & 45.9 & 23.3 & 1.97 \\ \hline
Superrefining & 19.6 & 1.9 & 61.1 & 32.5 & 1.88 \\ \hline
99.999+\% & 20.1 & 2.1 & 66.9 & 33.3 & 1.87 \\ \hline\hline
\end{tabular}
\end{center}
\vspace{-2mm}
\end{table}

\begin{figure}[!b]
\centerline{\includegraphics[width=0.44\textwidth]{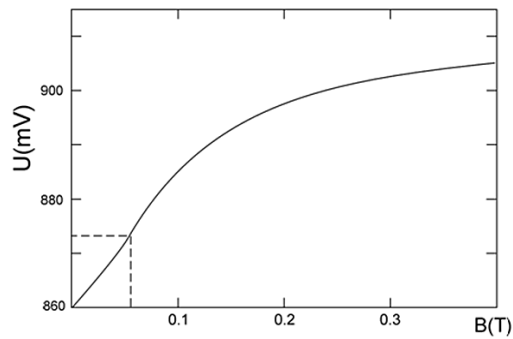}}
\caption{ Magnetic field dependence of transverse voltage for Ge-doped InSb. Dashed line indicates the flex point $B_{\text f} $ and the corresponding voltage $U(B_{\text f})$. } \label{fig2}
\end{figure}

As an example of a semiconductor, we consider germanium doped indium antimonide (InSb). In order to measure the transverse magnetoresistance of a rectangular parallelepiped-like sample ($0.25\times1.2\times5.4$~mm size), we place it into the pulse magnetic field, where the current $I = 1$~mA flew normally to the magnetic field lines. The transverse voltage contacts were soldered at a distance $\frac{1}{3 }l$ (where $ l $ is the length of the sample) from the current contacts. The magnetoresistance experimental data at $T=77$~K for this material are shown in figure~\ref{fig2}. The obtained value  coincides with the value for light holes mobility given by different authors \cite{Smi1}. The ratio of the corresponding conductivities is 0.07 which is in good agreement with the Hall-effect measurement results \cite{Uhr2}.

The most interesting example of applying the equations is calculating the minority charge carriers mobility of a superconductor in the critical temperature range.

It is known that high-temperature superconductors (cuprates) have a planar structure, and kinetic phenomena in these materials are connected with hybridized $\text{O}_{2p} \text{-} d_{x^2 \text{-} y^2}$ orbitals \cite{And}. Anderson \cite{And} describes it as a postulate (in the text --- ``dogma''). Until it is doped to 25\%, the Fermi surface is a simple hole surface around $X$ \cite{And}. Oxygen zone is hole-like and copper zone is electron-like (all the rest zones are located too far from Fermi level in order to take them into account), although there is known an opposite interpretation, when O-zone is electron-like and Cu-zone is hole-like \cite{Hir}. However, in the both cases these two zones provide two types of free charge carriers, the majority of which are, as a rule, holes. This follows from numerous Hall-effect experiments \cite{Poo}.

\begin{figure}[!t]
\centerline{\includegraphics[width=0.5\textwidth]{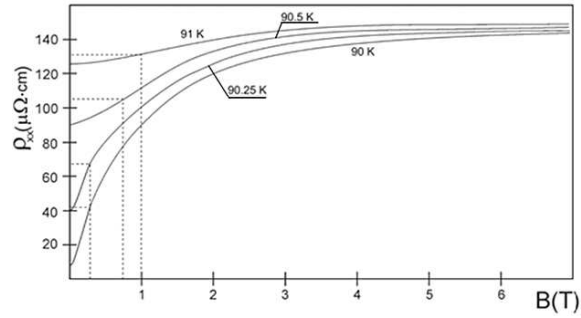}}
\caption{Magnetic field dependences of transverse resistivities for a multilayer superconductor $[\text{YBa}_{2} \text{Cu}_{3} \text{O}_{7}(72~\text{\AA})/ \text{PrBa}_{2} \text{Cu}_{3} \text{O}_{7} (12~\text{\AA})]_{25} $ at different temperatures near the critical one ($T_{\text{c}}=89.5$~K). Magnetic field is perpendicular to the layers. Dashed lines indicate flex points $B_{\text f} $ and resistivities in these points. Experimental data are obtained from temperature dependences of transverse resistivity at different magnetic fields from \cite{Qiu}. } \label{fig3}
\vspace{-1mm}
\end{figure}

Figure~\ref{fig3} shows the experimental dependences of transverse resistivity on the magnetic field inductance for several temperatures for high temperature multilayer superconductor $[\text{YBa}_{2} \text{Cu}_{3} \text{O}_{7}(72~\text{\AA})/$ $\text{PrBa}_{2} \text{Cu}_{3} \text{O}_{7} (12~\text{\AA})]_{25} $ near the critical point $T_{\text{c}}=89.5$~K, namely for $T>T_{\text{c}} $. The fact of the appearance of magnetoresistance at $T=91$~K testifies to the rise of a new sort of charge carriers. Indeed, the transverse resistance does not depend on magnetic field in the materials with one type of charge carriers as it can be seen from equation~(\ref{GrindEQ__2_}) substituting there $p=0$. Having determined for each curve the flex point $B_{\text f} $, zero field resistivity $\rho _{0} $, flex point resistivity $\rho _{\text f} $ and high field limit resistivity $\rho _{\infty } $, the basic parameters of the minority charge carriers can be calculated. The results of the calculations are shown in table~\ref{tab2} demonstrating a sharp increase of the negative sign charge carriers mobility, when the temperature approaches the superconductivity transition, which causes the same rate of their conductivity increase. At the same time, the majority charge carriers conductivity remains constant, which certainly shows their basic parameters to be constant. This abrupt conductivity change from hole to electron-like induces us to suppose that the electrons being minority charge carriers in the normal state play a majority role in the superconductive state.

Since one can find out a similar magnetoresistivity behavior near $T_{\text{c}} $ in other cuprate high temperature superconductors, doped with Nd, Tm, Bi, Ca, Sr, as well as in conventional ones \cite{Qiu,Hag1,Gal,Bre,Lat,For,Gin,Pui,Van,Usu,Wei,Not,Hag2,Smi2}, we can suppose that this temperature dependence of minority to majority charge carriers conductivity ratio  is general, including the case when the minority charge carriers are  positive, as it is for Nd-Ce-Cu-O \cite{Hag1}.

\begin{table}[!t]
\begin{center}
\caption{\label{tab2} Temperature dependence of charge carriers basic parameters in a layered superconductor $[\text{YBa}_{2} \text{Cu}_{3} \text{O}_{7}(72~\text{\AA})/ \text{PrBa}_{2} \text{Cu}_{3} \text{O}_{7} (12~\text{\AA})]_{25} $. The meanings of the symbols are the same as in table~\ref{tab1}.}
\vspace{2ex}{\footnotesize 
\begin{tabular} {|c|c|c|c|c|c|c|c|c|c|} \hline\hline
$T$, K & $\mu _{n}$, $\frac{\text{m}^{2\strut} }{\text{V}\cdot \text{s}}$ & $\mu _{p}$, $\frac{\text{m}^{2} }{\text{V}\cdot \text{s}}$ & $p$, $10^{27}~\text{m}^{-3}$ & $\frac{\sigma _{n}}{\sigma _{p}} $ & $\sigma _{p}$, $10^{5}~\text{\textohm} \cdot \text{m}$ & $\sigma _{n}$, $10^{5}~\text{\textohm} \cdot \text{m}$ & $n$, $10^{24}~\text{m}^{-3}$ & $a$ & $b$ \\ \hline\hline
91 & 0.7 & 0.0018 & 2.2 & 0.23 & 6.5 & 1.5 & 1.3 & 0.0059 & 390 \\ \hline
90.5 & 1.3 & 0.0018 & 2.2 & 0.67 & 6.5 & 4.4 & 2.1 & 0.0095 & 720 \\ \hline
90.25 & 7.1 & 0.0018 & 2.2 & 2.7 & 6.5 & 17.6 & 1.5 & 0.0068 & 3900 \\ \hline
90 & 40.9 & 0.0018 & 2.2 & 16.5 & 6.5 & 107 & 1.7 & 0.0077 & 22000 \\ \hline\hline
\end{tabular}}
\end{center}
\vspace{-2mm}
\end{table}

Note also that the application of our equations, that originate from semiclassic analysis of galvanomagnetic phenomena, is reasonable for this case since experimental curves shown  in figure~\ref{fig3} are in good agreement with the test equation~\eqref{GrindEQ__14_}.

The introduced interpretation of magnetoresistance behavior in superconductors also solves the problem of the Hall-effect anomaly near the critical temperature \cite{Qiu,Hag1,Gal,Bre,For,Gin,Pui,Van,Usu,Wei,Not,Hag2,Smi2}  which consists in the sign change of the Hall-effect in a low magnetic field at the temperatures approaching $T_{\text{c}} $ from the high temperature region. This phenomenon is common for both conventional \cite{Van,Usu,Wei,Not,Hag2,Smi2} and high temperature \cite{Qiu,Hag1,Gal,For,Gin,Pui,Hag2} superconductors. The majority of the authors explain it as the vortex motion concept, some authors are disposed to the pining influence, the others suppose this phenomenon to be connected with the change of electron to hole conductivity ratio \cite{Gal}.

We explain this sign reversal by the great electron mobility obtained above. As we can see from the expression for the Hall constant in a weak magnetic field \cite{Smi1}

\begin{equation} \label{GrindEQ__15} R_{\text{H}} =\frac{1}{e} \frac{p-nb^{2}}{(p+nb)^{2} }\,,
\end{equation}
the sign of $R_{\text{H}} $ will reverse negative when $b^{2} >p/n $ which, in its turn, is provided by the great value of electron mobility $\mu _{n} $.

Moreover, the so-called ghost critical field appears in some superconductors, that is the Hall-effect maximum field near the critical temperature \cite{Bre}. The prospect of our further research is to show that the maximum could appear in the solids with two types of holes.

\section{Conclusions}

  We can conclude that the introduced equations for determination of the minority  charge carriers mobility can be applied to all solid materials (probably not only for solid ones) giving new opportunities for their studying. The most interesting result is found for a superconductor showing a rapid increase of minority charge carriers mobility when the temperature approaches  the critical one from the normal state temperature region. We suppose that this rapid increase makes minority charge carriers responsible for the appearance of a superconductive state.

\newpage

\ukrainianpart

\title{Визначення основних параметрів неосновних носіїв заряду в твердих тілах на основі магнетоопору}

\author{Ю.О. Угрин\refaddr{label1}, Р.М.~Пелещак\refaddr{label1}, В.Б. Британ\refaddr{label1}, А.О. Вельченко\refaddr{label2}}
\addresses{
\addr{label1} Дрогобицький державний педагогічний університет ім. Івана Франка,\\
вул. І. Франка, 24, 82100 Дрогобич, Україна
\addr{label2} Білоруський державний аграрний технічний університет, \\ просп. Незалежності, 99, 220023 Мінськ, Білорусь
}

\makeukrtitle

\begin{abstract}
\tolerance=3000%
Запропоновано спосіб використання магнетоопору, як інструмент для  визначення основних параметрів носіїв заряду та відношення провідності неосновних носіїв заряду до основних в твердих тілах на основі аналізу кривої магнетоопору в рамках феноменологічної двозонної моделі.
Встановлено критерій застосовності цієї моделі.
В ролі прикладів застосування отриманих рівнянь приведено провідник, напівпровідник та надпровідник. Зі знайдених температурних залежностей згаданих вище величин в надпровідниках зроблено припущення про вирішальну роль неосновних носіїв заряду у виникненні надпровідного стану.
\keywords  провідник, напівпровідник, магнетоопір, Голл-ефект
\end{abstract}


\begin{thebibliography}{99}
\bibitem{Kuc}
   Kuchis E.V., Galvanomagnetic Effects and Methods of their Research, Radio i Sviaz’, Moscow, 1990, (in Russian).

\bibitem{Uhr1}
	Uhryn Y.O., Invention Patent No.~87695, Bulletin No.~15, 10.08.2009, (in Ukrainian).

\bibitem{Smi1}
    Smith R.A., Semiconductors, Cambridge University Press, New York, 1978.

\bibitem{Luc}
	L\"uck R., Phys. Status Solidi B, 1966, \textbf{18}, 49,
	\bibdoi{10.1002/pssb.19660180104}.

\bibitem{Luc1}
	L\"uck R., Doctoral Dissertation, Technische Hochschule, Stuttgart, 1965.

\bibitem{Ash}
	Ashcroft N.W., Mermin N.D., Solid State Physics, Cengage Learning, New Delhi, 2011.


\bibitem{Uhr2}
    Uhryn Y.O.,  Sheregii E.M., Fiz. Tekh. Poluprovodn., 1988, \textbf{22}, 1375, (in Russian).

\bibitem{And}
    Anderson P.W., The Theory of Superconductivity in the High-$T_\text{c}$ Cuprates, Princeton University Press, New Jersey, 1997.

\bibitem{Hir}
    Hirsch J.E., Marsiglio F., Phys. Rev. B, 1991, \textbf{43}, 424, 
     \bibdoi{10.1103/PhysRevB.43.424}.

\bibitem{Poo}
    Poole C.P. (Jr.), Farach H.A., Creswick R.J., Prozorov R., Superconductivity, Elsevier, Amsterdam, 2014.

\bibitem{Qiu}
    Qiu X.G., Jakob G., Moshchalkov V.V., Bruynseraede Y., Adrian H., Phys. Rev. B,
    1995, \textbf{52}, 12994,
     \bibdoi{10.1103/PhysRevB.52.12994}.

\bibitem{Hag1}
  	Hagen S.J., Smith A.W., Rajeswari M., Peng J.L., Li Z.Y., Greene R.L., Mao S.N.,
  	 Xi X.X., Bhattacharya S.,  Li~Q., Lobb C.J., Phys. Rev. B, 1993, \textbf{47}, 1064,
  	  \bibdoi{10.1103/PhysRevB.47.1064}.

\bibitem{Gal}
    Galffy M., Zirngiebl E., Solid State Commun., 1988, \textbf{68}, 929,
     \bibdoi{10.1016/0038-1098(88)90136-6}.

\bibitem{Bre}
    Breznay N.P., Kapitulnik A., Phys. Rev. B, 2013, \textbf{88}, 104510, 
       \bibdoi{10.1103/PhysRevB.88.104510}.

\bibitem{Lat}
    Latimer M.L., Berdiyorov G.R., Xiao Z.L., Peeters F.M., Kwok W.K., Phys. Rev. Lett., 2013, \textbf{111}, 067001,
     \bibdoi{10.1103/PhysRevLett.111.067001}.

\bibitem{For}
	Forr\'o L., Hamzi\'c A., Solid State Commun., 1989, \textbf{71}, 1099,
	 \bibdoi{10.1016/0038-1098(89)90719-9}.

\bibitem{Gin}
	Ginsberg D.M., Manson J.T., Phys. Rev. B, 1995, \textbf{51}, 515,
     \bibdoi{10.1103/PhysRevB.51.515}.
    
\bibitem{Pui}
	Puica I.,  Lang W.,  G\"ob W.,  Sobolewski R., Phys. Rev. B, 2004, \textbf{69}, 104513,
	  \bibdoi{10.1103/PhysRevB.69.104513}.

\bibitem{Van}
    Van Beelen H., Van Braam Houckgeest J.P., Thomas M.H.M., Stolk C., De Bruyn Ouboter R., Physica, 1967, \textbf{36},  241,
     \bibdoi{10.1016/0031-8914(67)90247-9}.

\bibitem{Usu}
    Usui N., Ogasawara T., Yasuk\=ochi K., Phys. Lett. A, 1968, \textbf{27}, 139,
     \bibdoi{10.1016/0375-9601(68)91169-9}.

\bibitem{Wei}
    Weijesenfeld C.H., Phys. Lett. A, 1968, \textbf{28}, 362,
     \bibdoi{10.1016/0375-9601(68)90338-1}.

\bibitem{Not}
  	Noto K.,  Shinzawa S., Muto Y., Solid State Commun., 1976, \textbf{18}, 1081,
  	 \bibdoi{10.1016/0038-1098(76)91245-X}.

\bibitem{Hag2}
    Hagen S.J., Lobb C.J., Greene R.L., Forrester M.G., Kang J.H., Phys. Rev. B, 1990, \textbf{41}, 11630, \\
     \bibdoi{10.1103/PhysRevB.41.11630}.

\bibitem{Smi2}
    Smith A.W., Clinton T.W., Tsuei C.C., Lobb C.J., Phys. Rev. B, 1994, \textbf{49}, 12927, \\
     \bibdoi{10.1103/PhysRevB.49.12927}.

\end{thebibliography}
\end{document}